\documentstyle[twocolumn,eqsecnum,aps]{revtex}
\begin{document}
\draft
\title{Mesons a in Collinear QCD Model}
\author{M. Burkardt}
\address{Department of Physics\\
New Mexico State University\\
Las Cruces, NM 88003-0001\\U.S.A.}
\maketitle
\begin{abstract}
A phenomenological model for the quark structure of mesons is considered. 
The model is based on the tube model for QCD, where all quanta with nonzero 
transverse momenta are neglected. In the limit that the mass term of the 
gluons goes to infinity, the model is equivalent to a combination of the 
't Hooft and Gross-Neveu models and can be solved semi-analytically.
The model has the properties of confinement, chiral symmetry breaking
and asymptotic freedom and thus resembles QCD in three key respects.
Spectra, distribution amplitudes and form factors of mesons are analyzed.
\end{abstract}
\narrowtext
\section{Introduction}
Many high energy scattering processes, probe the quark-gluon 
structure of hadrons predominantly in one direction \cite{dgr:elfe}. It is 
thus suggestive to consider a phenomenological model for QCD, where the
transverse momenta of all constituents are neglected. At a semi-classical
level, one can think of this approximation as a formulation of QCD in
a ``tube'' with periodic boundary conditions. By taking the radius
of the tube very small, all modes other than ${\vec k}_\perp=0$ have
a very high energy and thus freeze out. The model which will be
considered in this work is based on such a classical reduction of the
QCD Lagrangian to an effective 1+1 dimensional theory, which is
subsequently quantized. It should be emphasized that the model
thus obtained is not a dimensional reduction at the level of the
full quantum theory, which is slightly more subtle. As a consequence,
the model should not be considered a rigorous approximation to full
QCD but rather a phenomenological model. Furthermore, it should be
noted that even though the model is formulated in 1+1 dimensions,
one should not consider it a mere toy model but rather a simple
(because all ${\vec k}_\perp$ vanish) phenomenological model for
$QCD_{3+1}$. 

While it may be possible to understand many features of high energy
scattering experiments in terms of a simplified effective theory,
which is reduced to degrees of freedom with vanishing transverse 
momenta, boost invariance often seems essential for microscopic descriptions
of such experiments. Furthermore, high energy scattering experiments typically
probe correlations along light-like directions \cite{dgr:elfe}. 
It thus makes sense to employ light-front (LF) quantization 
\cite{world,osu:all,brazil,mb:adv} when studying dimensionally reduced QCD. 

Such a model has been considered in Ref. \cite{tube} for glueballs and
in Ref. \cite{anton} for mesons, where discrete light-cone quantization 
(DLCQ) \cite{pa:dlcq} has been used as a numerical tool.
However, due to the rather singular behavior
of the fermion wave functions, the resulting spectra are not
very well converged numerically when the quark masses become small. 
For form factors, which we would
like to consider in this work, it is even more important to describe
the end-point behavior of the wave functions very accurately.
Improvement methods based on DLCQ \cite{brett:imp} have so far
only been developed for gluons.

In order to develop a model which can be solved semi-analytically,
we will consider a modified version of the model in Ref. \cite{anton}
where we assume that the mass term for the effective gluon is
very large and thus gluon degrees of freedom can be eliminated
perturbatively. The resulting effective interaction for the quarks
resembles 't Hooft's large-$N_c$ QCD in 1+1 dimensions \cite{thooft}
with an additional helicity dependent Gross-Neveu interaction
\cite{gn,brian}

Because the matter degrees of freedom in collinear QCD couple to
a 1+1 dimensional gauge theory, confinement is an almost trivial
feature of the model.
The effective Gross-Neveu interaction, which arises from eliminating 
the transverse gluons, provides an induced mass for the quarks and
thus leads to spontaneous breaking of chiral symmetry. Furthermore,
the Gross-Neveu interaction is also asymptotically free.
Thus, even though collinear QCD cannot be derived rigorously as an 
approximation to full QCD, the mere fact that it shares with real QCD
the important properties of confinement, chiral symmetry breaking as 
well as asymptotic freedom makes it worth while to investigate
collinear QCD as a phenomenological model.

As a first application of the model, we will study the low lying meson 
spectrum. This part of the calculation will also be used to determine 
the free parameters of the model. Once all terms in the Hamiltonian are
fixed, we will then proceed to  calculate other observables, specifically
distribution amplitudes (i.e.
light-cone wave functions) and form factors.

\section{The Collinear QCD Model for Mesons}
The basic idea of collinear QCD (also referred to as the tube model) 
is to start from the classical QCD Lagrangian and to neglect all 
dependences on transverse coordinates (transverse with respect to an 
arbitrarily chosen, but fixed direction), yielding
\begin{eqnarray}
{\cal L}&=& \bar{\psi} \left[ \gamma^+ \left(i\partial_+-
gA_+\right)+\gamma^- \left(i\partial_--
gA_-\right)-g{\vec \gamma}_\perp {\vec A}_\perp -m\right]\psi
\nonumber\\
&-& \frac{1}{2}\mbox{tr} G^{\mu \nu}G_{\mu \nu}
-\frac{\lambda^2}{2}{\vec A}_\perp^2 ,
\label{eq:lager}
\end{eqnarray}
where we have introduced LF coordinates 
$A_\mp=A^\pm =\left(A^0\pm A^3\right)$. To keep the calculations simple, which will help in
identifying the essential physics,
only the limit $N_c\rightarrow \infty$
will be considered here.

Neglecting modes with ${\vec k}_\perp \neq 0$
breaks invariance under ${\vec x}_\perp$ dependent gauge
transformations. Therefore a mass for the transverse component of
the gauge field has been added, since it is now no longer
protected by gauge invariance. Eq. (\ref{eq:lager}) is still
invariant under gauge transformations that depend on
$x^\pm$ only and we are thus free to choose the gauge $A^+=0$.
\footnote{Explicit zero-modes, i.e. modes that do not depend on 
$x^-$, will be neglected throughout this work.}.
Neither $A^-$ nor $\psi_-$ 
(where $\psi_\pm \equiv \frac{1}{2}\gamma^\mp \gamma^\pm \psi$)
are dynamical degrees of freedom, since 
their LF-time ($x^+$) derivative does not enter the Lagrangian
(\ref{eq:lager}). This degrees of freedom are eliminated from
the Lagrangian [Eq. (\ref{eq:lager})] using the corresponding
constraint equations (Euler Lagrange equations) before
quantizing the theory.

The dynamical degrees of freedom of the model are $\psi_+$
and ${\vec A}_\perp$, which are quantized canonically. 
In summary, what the dimensional reduction yields is
a 1+1 dimensional gauge theory, coupled to both scalar
matter in the adjoint representation (${\vec A}_\perp$) and
fermionic matter in the fundamental representation ($\psi $).
Both ${\vec A}_\perp$ and $\psi $ have internal helicity degrees
of freedom --- remnants of the underlying 3+1 dimensional
theory. 

In addition to the coupling of the matter fields to the 1+1 dimensional
gauge theory, there is also a direct, Yukawa-like, coupling between
fermions and ${\vec A}_\perp$, which flips the helicity of the fermions.

Note that the dimensional reduction procedure, the choice of gauge ($A^+=0$) as 
well as the quantization plane ($x^+=0$) are all manifestly invariant 
under rotations about the $z$-axis.
As a result, the $z$-component of the total angular momentum is a
conserved quantity in the model. Since there is no orbital angular momentum
left after the dimensional reduction, this means that the total
z-component of the spin, i.e. the sums of z-components of the spins of quarks, 
plus anti-quarks plus gluons, is conserved. This feature of the model will 
be very helpful when it comes to classifying states.

Such a model has been studied in Ref. \cite{anton}, where approximate 
numerical solutions have been obtained using a Fock space expansion 
and DLCQ \cite{pa:dlcq}. In Ref. \cite{anton} , one can also find an explicit
expression for the dimensionally reduced LF Hamiltonian for QCD, which
will not be reproduced here in its full generality, since we will 
consider only a solvable limiting case of the model in this work.
 
By allowing the mass term for ${\vec A}_\perp$ to become arbitrarily 
large (with appropriate rescaling of the coupling constant in order
to keep the physics nontrivial), it is possible to systematically
eliminate ${\vec A}_\perp$, yielding an effective interaction
which acts only on the fermion degrees of freedom. For this we note
that, as long as they are not diverging, all interactions
of ${\vec A}_\perp$ become negligible in this limit. 
Furthermore, we can neglect all interactions in intermediate states
which contain quanta of ${\vec A}_\perp$ and are thus highly
off-shell. This makes
it an easy task to eliminate ${\vec A}_\perp$ using a LF Tamm Dancoff
procedure. First this gives rise to a kinetic mass counter-term
for the fermions. As is explained in the Appendix, the Tamm-Dancoff
procedure also requires to renormalize the one gluon vertex.
After eliminating all gluon dressing from the quark lines, we
are thus left with a fermion field that has a mass term $M$.
\footnote{Strictly speaking we would have to keep separate
kinetic and vertex mass terms. However, as is discussed in  
Appendix \ref{app:vertex},
choosing them equal gives rise to a consistent solution.} 

\begin{figure}
\unitlength1.cm
\begin{picture}(15,6)(1,1)
\includegraphics{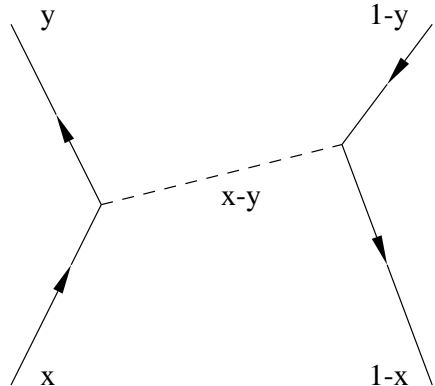}
\end{picture}
\caption{Scattering of a quark with initial LF-momentum $x$ from
an anti-quark of initial momentum $1-x$. The dotted line represents 
the exchanged gluon. Only one time ordering is shown.
}
\label{fig:scatt}
\end{figure}

In order to determine the effective $q\bar{q}$ interaction arising
from eliminating ${\vec A}_\perp$, let us consider 
$q\bar{q}$ scattering (Fig. \ref{fig:scatt}) \footnote{Once again,
since $\lambda \rightarrow \infty$, all other interactions can be
neglected in intermediate states with ${\vec A}_\perp$ quanta.}
Since the quark helicity flips at the quark gluon vertices and since
total (quark+gluon) helicity is conserved at each vertex,
the one gluon exchange interaction vanishes identically for helicity $\pm 1$
mesons, i.e. when quark and antiquark have the same helicity.
Thus, as our first result, we find that mesons with helicity $\pm 1$ are
in our model described by 't Hooft's equation only 
\begin{equation}
\tilde{\mu}_n^2 \psi(x) = \left( \frac{M^2}{x} +\frac{M^2}{1-x}\right)
\psi (x) +G^2\int_0^1 dy \frac{ \psi (x)-\psi(y)}{(x-y)^2} .
\label{eq:thooft}
\end{equation}
The only role played by the transverse gluons for these mesons
is to give the quarks a ``constituent mass'' $M$.

The situation is different for mesons with helicity zero, i.e. when
quark and antiquark have opposite helicity, in which case the
one gluon exchange matrix element is nonzero and reads 
for the time ordering depicted in Fig. \ref{fig:scatt} \footnote{
The overall coupling is irrelevant at this point, since it
is renormalized anyways.}
\begin{eqnarray}
\langle x |V_{oge}^{Fig. 1} |y\rangle &\propto&
\frac{g^2\frac{\Theta(x-y)}{x-y}\left( \frac{1}{x}-\frac{1}{y} \right)\left( \frac{1}{1-x}-\frac{1}{1-y} \right)}{2P^+P^-- \frac{M^2}{y} -\frac{M^2}{1-x} - \frac{\lambda^2}{x-y}}
\nonumber\\
&\stackrel{\lambda^2\rightarrow \infty}{\longrightarrow} &
\frac{-g^2\Theta(x-y)}{\lambda^2}\left( \frac{1}{x}-\frac{1}{y} \right)\left( \frac{1}{1-x}-\frac{1}{1-y} \right) .
\nonumber\\
& &
\label{eq:voge}
\end{eqnarray}
The other time-ordering gives 
(for $\lambda \rightarrow \infty$)
the same expression, but with $\Theta(y-x)$, so that
the $\Theta$-function disappears in the sum of the two time-orderings
\begin{eqnarray}
\langle x |V_{oge} |y\rangle \propto \frac{-g^2}{2\lambda^2}
& &\left[\left( \frac{1}{x}+\frac{1}{1-x}\right)\left( \frac{1}{y}+\frac{1}{1-y}\right)
\right. \nonumber\\
& &\quad +\left.\left( \frac{1}{x}-\frac{1}{1-x}\right)\left( \frac{1}{y}-\frac{1}{1-y}\right)
\right. \nonumber\\
& &\quad +\left.  \frac{2}{x(1-x)}
+  \frac{2}{y(1-y)}\right],
\label{eq:voge1}
\end{eqnarray}
where the different Lorentz structures appearing in Eq. (\ref{eq:voge}) have 
been separated. Let us first consider the ``pion'' channel, 
i.e. mesons with helicity
zero, an odd spin wave function and an even orbital wave function for the
quarks. In this channel, the effective interaction reads
\begin{eqnarray}
\langle x |V^{eff,\pi}_{oge} |y\rangle \propto \frac{-g_{eff}^2}{2}& &\left[\left( \frac{1}{x}+\frac{1}{1-x}\right)\left( \frac{1}{y}+\frac{1}{1-y}\right)
\right. \nonumber\\
& &\quad +\left.  \frac{2}{x(1-x)}
+  \frac{2}{y(1-y)}\right] 
\label{eq:voge2}
\end{eqnarray}
where $g_{eff}^2 = g^2/\lambda^2$
The first interaction term in Eq. (\ref{eq:voge2}) is the same as what one
gets in the chiral Gross-Neveu model, i.e. an attractive s-channel
$\bar{\psi}_ii\gamma_5\psi_i \bar{\psi}_ji\gamma_5\psi_j$ interaction,
where $i,j$ are color indices.
Such an interaction leads to a divergence when it is iterated and one has to
introduce a cutoff. From studies of the Gross-Neveu model \cite{gn,brian},
it is known that the the coupling $g_{eff}$ goes to zero as the cutoff is sent to
infinity (asymptotic freedom!). 
The other two terms in Eq. (\ref{eq:voge2}) are what one gets if one
introduces an s-channel vector interaction of the form
$\bar{\psi}_i\gamma_\mu\psi_i \bar{\psi}_j\gamma^\mu\psi_j$ in $1+1$ 
dimensions.
Such an interaction does not lead to divergences (in covariant perturbation
theory) --- neither when iterated with itself nor in conjunction
with the $QCD_{1+1}$ or Gross-Neveu interactions.
Thus as the running coupling goes to zero, this s-channel vector interaction
becomes less and less important, until it disappears when the cutoff goes to
infinity. In the effective interaction, one can thus omit this term entirely,
yielding
\begin{equation}
\langle x |V^{eff,\pi}_{oge} |y\rangle \propto \frac{-g_{eff}^2}{2}\left( \frac{1}{x}+\frac{1}{1-x}\right)\left( \frac{1}{y}+\frac{1}{1-y}\right) .
\label{eq:voge2b}
\end{equation}
In the scalar channel (total helicity zero, spin part odd, orbital part odd)
one finds the same result as for the chiral Gross-Neveu model in the scalar channel
\begin{equation}
\langle x |V^{eff,\sigma}_{oge} |y\rangle \propto \frac{-g_{eff}^2}{2}\left( \frac{1}{x}-\frac{1}{1-x}\right)\left( \frac{1}{y}-\frac{1}{1-y}\right)
\end{equation}
and also the relative sign and strength between the $\pi$ and $\sigma$ channel
are consistent with the chiral Gross-Neveu model.

In addition to this contact interaction from eliminating the transverse
Gluon field, the mesons are also governed by the $QCD_{1+1}$ interactions.
Thus, after all the dust has settled, we are left with an effective
two dimensional field theory, describing mesons 
both in the $\pi$ and $\sigma$ channels, which in covariant notation
reads \footnote{One can arrive at the same result also by means of a 
rather lengthy Fierz rearrangement. However, it is much more illuminating
to derive this result in the LF helicity basis as has been done above.}
\begin{eqnarray}
{\cal L} &=& \bar{\psi}\left( i\partial \!\!\!\!\!\!\not\,\; -
\frac{G}{\sqrt{N}}A\!\!\!\!\!\!\not\;\,-m\right)\psi
- \frac{1}{2}\mbox{tr}\left(F_{\mu \nu}F^{\mu \nu}\right)
\nonumber\\
& &-\frac{\gamma}{2N}\left[ \left(\bar{\psi}\psi\right)^2
-\left(\bar{\psi}i\gamma_5\psi\right)^2\right],
\label{eq:hgn0}
\end{eqnarray}
where $\psi$ is a two-spinor with $N$ color components 
(fundamental representation),
$F_{\mu \nu}=\partial_\mu A_\nu
-\partial_\nu A_\mu +i\frac{G}{\sqrt{N}} \left[A_\mu,A_\nu\right]$ 
is the gauge field and the limit $N\rightarrow \infty$ is implied.
Obviously, for $\gamma =0$ one obtains the 't Hooft model
($QCD_{1+1}$ in the limit $N\rightarrow \infty$) \cite{thooft}, 

Finally, let us turn to mesons with zero total helicity, but
even helicity wave functions. By the same reasoning as above,
one obtains the same interaction as in the $\pi-\sigma$ channels,
but with the sign reversed (because the interaction is a
spin exchange interaction), i.e. they act repulsively.
In non-perturbative calculations, there is a crucial difference between
renormalized zero-range interactions that are attractive and those that are
repulsive! This fact is best illustrated in the context of a simple
example, the non-relativistic Schr\"odinger equation in two
spatial dimensions with a delta function interaction. Since a
delta function in two dimensions leads to a logarithmic divergence,
one needs to introduce a cutoff. Suppose now that the delta function 
is attractive. Then, in order to keep a finite mass of the bound state,
the coupling constant must go to zero as the cutoff goes to infinity
--- otherwise the wave-function gets sucked into the singularity.
Now suppose one calculates the effect of a repulsive interaction with the
same strength but opposite (repulsive) sign. In this case, there is
no enhancement of the interaction due to being sucked in and as a 
consequence the repulsive effects of the interaction completely disappear 
as the cutoff is sent to infinity.

The same happens here with the spin-even helicity zero interaction.
Because of the symmetries of the underlying action, the coupling 
constant is the same as in the attractive case ($\pi-\sigma$ channel)
but of opposite (repulsive) sign. Since non-perturbative
renormalization for the $\pi-\sigma$ channel (as done in Appendix \ref{sec:hgn})  
yields a vanishing bare coupling, and since the same bare coupling
acts in the repulsive case, one finds that the effective interaction
from one gluon exchange in the repulsive channels completely
disappears. The spin-even helicity 0 mesons are thus described
by 't Hooft's equation (without Gross-Neveu interaction).
This is a very interesting result, since it implies that the
spin-even helicity 0 mesons are described by the same
bound-state equation as the helicity $\pm 1$ mesons, and they thus
form degenerate triplets --- just as one would expect from
vector and axial vector mesons!

In summary, after separating the spin part of the effective interaction
induced by the transverse gluons, one finds that scalar and pseudo-scalar
mesons are described by the 't Hooft -- Gross -- Neveu interaction 
[Eq. (\ref{eq:hgn0})], while vector and axial vector mesons are described
by 't Hooft's equation [Eq. (\ref{eq:hgn0}), but with $\gamma =0$].
\footnote{However, the effective masses of the quarks in the vector and axial
vector channels are the same as the ones generated by the Gross-Neveu
interactions in the scalar and pseudo-scalar channels.} 
This is one of the principal results of this paper.

Quantum numbers are associated with these states by considering vacuum
to meson matrix elements of operators with definite quantum numbers
(such as $\bar{\psi}\psi$ and $\bar{\psi}i\gamma_5\psi$).
The resulting quantum numbers are consistent with the
quantum numbers that one can intuitively guess by merely considering
the degeneracy of states and their naive quark model parity properties.
Using these rules, one obtains the assignments of quantum numbers
shown in Table I. Of course, while the above assignment of quantum numbers
may appear reasonable, it is still to some extent arbitrary, since the
model has no rotational symmetry. One should always keep this fact in mind
when comparing the model results to the experimentally measured meson spectra.

Note that, in the limit of infinite mass for the $A_\perp$ field considered 
here, the model supports only states with helicity up to $h=1$. 
Tensor and higher spin mesons are absent in this limit of the model.
\begin{table}
\label{tab:jp}
\begin{tabular}{|c|c|c|c|}
helicity & hel. symmetry & orbital symmetry & $J^P$ \quad\quad\\ \hline
0 & odd & odd & $0^+$ \quad\quad\\
0 & odd & even & $0^-$ \quad\quad\\
0 & even & odd & $1^+$ \quad\quad\\
1 & even & odd & $1^+$ \quad\quad\\
0 & even & even & $1^-$ \quad\quad\\
1 & even & even & $1^-$ \quad\quad\\ 
\end{tabular}
\caption{Assignment of $J^P$ quantum numbers to the states of the model. One 
of the remarkable results of this model is that, despite the rotationally 
non-invariant formulation, states with $h=0$ and those with $h=\pm 1$ are 
exactly degenerate (this is true for both vector and axial vector mesons).
}
\end{table}
\section{Numerical Results}
\subsection{Meson Spectrum}
In the case of vector and axial vector mesons, the numerical
calculations amount to just solving 't Hooft's equation,
which has been well studied and nothing further needs to be
said about this part. For the scalar and pseudo-scalar mesons,
which are described by a combination of the 't Hooft and Gross-Neveu
models, the calculational methods to obtain spectra and wave functions
are described in the Appendix \ref{sec:hgn}. In both cases (i.e.
with and without Gross-Neveu interaction) the numerical calculations
were based on variational calculations with a basis of Jacobi
polynomials, such that the correct end-point behavior of the wave
functions was manifestly built in. typically, a basis size of 5-10
states was already sufficient to give the lowest eigenvalues
with at least four significant digits.

The model contains three free parameters: the $QCD_{1+1}$ coupling,
the Gross-Neveu coupling and the effective (constituent) quark mass.
These parameters were fixed by using as input the 
masses of the $\pi$, $\rho$ and $\rho (1450)$ mesons.
The masses of both the $\rho$ and the $\rho (1450)$ do not
depend on the Gross-Neveu coupling, and they thus determine
the values of the bare parameters in the $QCD_{1+1}$ interactions,
yielding $m_q = 288 MeV$ and $\sqrt{G^2/\pi}= 278 MeV$ respectively.
\footnote{Note that this quark mass includes already the mass 
renormalization due to the Gross-Neveu interactions and should thus 
be regarded as a constituent quark mass.}
The value of the renormalized Gross-Neveu coupling is then obtained 
by performing a subtraction at the $\pi$ pole
as explained in Appendix \ref{sec:hgn}.

The physical string tension can be read off directly from the
static $Q\bar{Q}$ potential in $QCD_{1+1}$
\begin{equation}
V(x) = \frac{G^2}{2}|x| .
\end{equation}
For the above model parameters, one thus obtains a physical string tension 
of $\sigma = (349MeV)^2$, which is slightly lower than the preferred 
value used by the lattice community [$\sigma_{latt} \approx (400MeV)^2$].

Having determined all free parameters, we can now turn to calculate
masses of other hadrons. When comparing with experimental data,
the comparison is always done with isospin one states. The reason for
this choice is that the model makes a large $N_c$ approximation, and
it is generally expected that such an approximation is better
for isospin one states than for isospin zero states, since
isospin zero states in QCD are affected by mixing with gluonic
components and such mixing is turned off in the large $N_c$ limit.
Nevertheless, identifying the model spectrum with isospin 1 states
is still to some extent an arbitrary choice.

Numerical results for the lightest mesons and their first radial excitations
are shown in Table II. Given the simplicity of the model,
the agreement  with the experimental data \cite{pdg} is spectacular!

One may argue which of the mesons in Table II should be used as input parameters,
particularly since some of these states are not believed to be simple quark model
states but rather complicated ``molecule'' type states and one would not expect
that such mesons are adequately described by a large $N_C$ model.
However, since all mesons in Table II are fitted quite well, picking a different
set of input parameters has only a minor impact on the quality of the fit.

However, and this is more important, as we discussed in the previous session, 
there is the issue of
assignment of quantum numbers, which reflects itself here in picking states
from the particle data table \cite{pdg} and comparing the model results to
this data. Because of the truncation of transverse momenta, the model has less
degrees of freedom than QCD (or a 3+1 dimensional quark model). Therefore,
not all mesons with spin 0 or 1 have a model equivalent. By making 
(somewhat arbitrary) assignments of quantum numbers, one does in essence pick a
subset of mesons from the particle data table and fits these mesons.
As an example, the way we assigned quantum numbers, there are $a_0$ and $b_1$ mesons
in the model but no $a_1$ mesons. In principle, one might thus be tempted
to generate a fit to a different sub-set of mesons. However, since the above choice
of quantum numbers seemed to be by far the most reasonable choice, no such results
will be presented here. Nevertheless, one could always keep this in mind as an option.
\begin{table}
\label{tab:spectrum}
\begin{tabular}{|c|l|l|l|}
meson & $J^P\quad\quad$ & model (MeV) $\quad$& exp.(MeV)$\quad$\\ \hline
$\pi$ & $0^-$ & 139 (input) & 139 \\
$\rho$ & $1^-$ & 769 (input) & $769 \pm 1$ \\
$a_0$ & $0^+$ & $978$ & $984 \pm 1$\\
$b_1$ & $1^+$ & 1249 & $1231 \pm 10$ \\
$\pi (1300)$ & $0^-$ & $1339$  & $1300 \pm 100$\\
$\rho (1450)$ & $1^-$ & 1465 (input)  & $1465 \pm 25$ \\
$a_0^* $ & $0^+$ & $1606 \pm 2$ &  ? \\
$b_1^*$ & $1^+$ & 1707 & ?
\end{tabular}
\caption{Calculated masses of the lightest mesons and their 
first radial excitations for
each quantum number. Uncertainties are indicated
only when they exceeded 1 MeV.
}
\end{table}
Motivated by the unexpected success of the model for calculations of the
meson spectrum, we will in the following consider other observables as well.
Of course, because of the severe approximations used in constructing the model,
we will limit ourselves to observables where one would expect that the 
truncation of transverse momenta is not so critical. Furthermore, we will
focus on those observables where the LF formalism is helpful in simplifying
the calculations.

\subsection{Meson Wave Functions}
One quantity of great interest, which is easily accessible in our
model, is the so called (twist-2) light-cone wave function of mesons, defined
by a correlator along a light-like direction, i.e. for example
\begin{equation}
\psi_\pi (z) \propto \int d^2x dx^-
\langle 0 | \bar{\psi}(0,x_\perp) \gamma^+\gamma_5 \psi (x^-,x_\perp) |
\pi \rangle e^{i p^+x^-z}
\label{eq:psipi}
\end{equation}
in the case of the $\pi$, and correspondingly for the $\rho$.
In gauges other than the LF-gauge $A^+=0$, a gauge string in Eq. (\ref{eq:psipi})
is understood. LF time $x^+$ is the same for the two field operators.
Since the definition of these LF wave functions involves only longitudinal
correlations, it seems sensible to consider them in our collinear model.
\begin{figure}
\unitlength1.cm
\begin{picture}(15,7)(1,-7)
\includegraphics{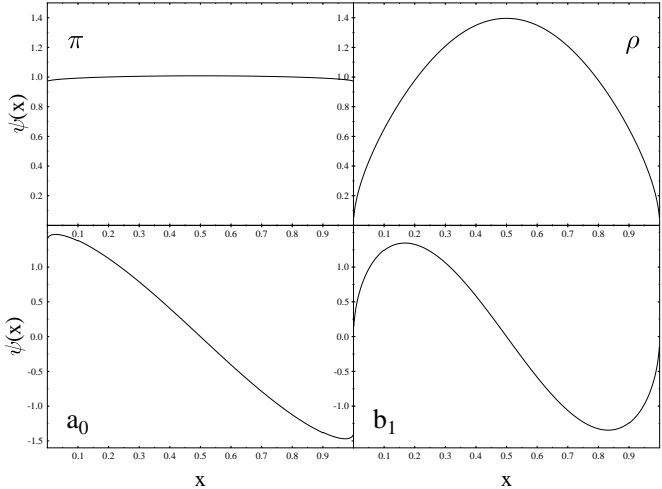}
\end{picture}
\caption{Light-cone wave functions for the 4 lightest mesons.}
\label{fig:psi}
\end{figure}
Numerical results for the LF wave functions of the four lightest mesons are 
shown in Fig. \ref{fig:psi}. The wave functions for the $\pi$ and $\rho$
are normalized to $f_\pi$ and $f_\rho$ respectively, since this is common
in discussions of these mesons. No such physically motivated normalization
exists for the $a_0$ and $b_1$ mesons and they are thus normalized to
$\int_0^1dx \psi^2(x)=1$. The $\rho$ and $b_1$ are just solutions to
`t Hooft's equation. The boundary behavior ($x\rightarrow 0,1$) is
$\psi \sim x^\beta \, [(1-x)^\beta]$, where
$\pi \beta \cot \pi \beta = 1-m_q^2\pi/G^2$ and the number of nodes in the
wave function increases with the excitation energy of the corresponding meson.

Obviously, this a crude model and one should be careful when comparing these
results with other calculations or with experimental data. In particular,
since modes with large transverse momenta have been omitted, one should
only compare to calculations or data at a low momentum scale.
With these caveats in mind, Fig. \ref{fig:psi} shows several interesting
features, which might be relevant for the real world. First of all, the
pion wave function is nearly flat and it does not vanish at the end-points.
This is not just a numerical coincidence, as one can see by studying vacuum
to meson matrix elements of the color singlet axial current
\begin{equation}
f_np^\mu = \langle 0|j^\mu_5 |n\rangle .
\end{equation}
In the chiral limit, $\partial_\mu j^\mu_5 =0$ and thus $f_n$ must vanish
--- unless $\mu_n^2=0$. Only massless mesons are allowed to have 
$f_n \neq 0$ in the chiral limit. In terms of the LF wave functions,
\begin{equation}
f_n = \sqrt{\frac{N_C}{\pi}} \int_0^1 dx \phi_n(x)  
\end{equation}
and thus (regardless of the quark mass)
\begin{equation}
\frac{\pi}{N_C} \sum_n f_n^2 =1.
\label{eq:sr}
\end{equation}
If the $\pi$ is the only massless meson in the chiral limit 
(it is in this model), then Eq. (\ref{eq:sr}) implies
\begin{equation}
\int_0^1 dx \phi_\pi(x)=1,  
\end{equation}
which, together with the normalization condition 
$\int_0^1 dx \phi_\pi^2(x)=1$ and the Cauchy-Schwarz inequality, implies
$\phi_\pi(x)=1$, i.e. a constant $\pi$ wave function.
It should be emphasized that this argument does not work in more
than 1+1 dimensions, where one also has to integrate over ${\vec k}_\perp$.
However, in any 1+1 dimensional model with chiral symmetry and with
only one massless meson state the wave function of this state must be flat.
This statement holds even in the presence of higher Fock components,
in which case one can also prove that (again if there is only one  
massless meson state) its wave function must be purely valence and flat.

The $\rho$ meson wave function is peaked in the middle and it does vanish at
the end points. The main reasons for the difference between the $\pi$ and
$\rho$ meson wave functions are the stronger binding and 
the attractive short range interaction, which acts only in the pion 
channel. In a non-relativistic Schr\"odinger picture, the main difference
between the $\pi$ and the $\rho$ is that the $\pi$ wave function gets 
completely ``sucked'' into the region of the attractive short range 
interaction, thus leading to very high momentum components its wave 
function. End-points of the wave function on the LF, correspond to
high momenta in an equal time picture, which explains why the 
$\pi$ wave function is much larger than the $\rho$ wave function near 
the kinematical end-points ($x=0$ and $x=1$). 

Scalar and axial vector mesons show similar features as the $\pi$ and
$\rho$ respectively, except that there is a node in the wave function,
which arises since these mesons are simply the second lowest states in infinite
ladders of states with an increasing number of nodes --- very much similar 
to non-relativistic states in one space dimension.
The reason for the difference between the end-point behavior of the
$a_0$ and $b_1$ wave functions is the same as for the $\pi$ and $\rho$.

Note that parton distribution functions in this model are simply the
squares of the distribution amplitudes in this model, both because 
higher Fock components are negligible and because there are no integrals
over transverse momenta. However, before comparing the resulting parton  
distributions with experimentally measured structure functions, one should
bear in mind that the collinear approximation, i.e. neglecting quanta with
${\vec k}_\perp \neq 0$, makes sense only at low scales. 
Since it was not clear how to determine the starting scale for QCD evolution
within the model, no attempt was made to compare the resulting
structure functions to the available experimental data.

\subsection{Meson Form Factors}
\label{sec:form}

\begin{figure}
\unitlength1.cm
\begin{picture}(15,7)(.9,-8)
\includegraphics{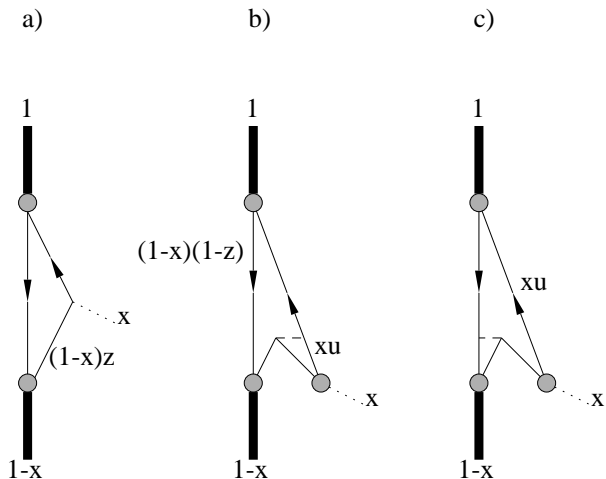}
\end{picture}
\caption{Time ordered diagrams contributing to the vector form factor of the 
$\rho$ meson: a) impulse term, b) vertex correction, c) exchange current.
The grey blobs represents the meson wave function and the dressed quark 
photon vertex.
}
\label{fig:ffrho}
\end{figure}

In the case of the $\rho$ meson, the 3 diagrams depicted in Fig. 
\ref{fig:ffrho} contribute to the elastic form factor.
As a side remark, it should be emphasized that in the LF framework,
one can usually calculate form factors directly by taking matrix elements
between the states and one does not need to worry about various
diagrams (at least for the ``good'' component $j^+$ of the current).
However, in the large $N_C$ limit, the bound state equations are greatly
simplified by leaving out all components of the wave function that are
$1/N_C$ suppressed. On the LF this not only eliminates non-planar
diagrams, but also diagrams which are planar but which contains
fermion lines that go forth and back in time. When one calculates
form factors, there are diagrams where the photon couples to such
backward going lines, which become suddenly allowed, even on the LF, 
as long as
the $+$ component of the momentum transfer is nonzero. Examples of such
diagrams are shown in Fig. \ref{fig:ffrho} b) and c).

\begin{figure}
\unitlength1.cm
\begin{picture}(15,7)(-0.3,-8)
\includegraphics{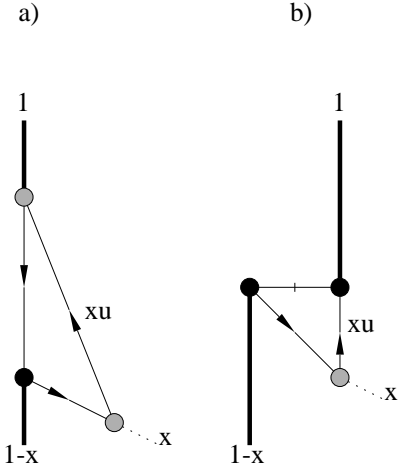}
\end{picture}
\caption{Additional time ordered diagrams contributing to the vector form 
factor of the $\pi$. The grey blobs represents the meson wave function
as the dressed quark photon vertex, and the black blob the coupling
of a quark to the meson via the Gross-Neveu interaction.
The slashed fermion line corresponds to an instantaneous fermion exchange
interaction.}
\label{fig:ffpi}
\end{figure}

The expression for the elastic form factor of the $\rho$ meson is identical
to the corresponding expression for mesons in $QCD_2$ \cite{einhorn}. The
reason is that the Gross-Neveu interaction neither contributes to the
equation for the $\rho$ nor does it contribute to the interaction in the
photon (vector current!) channel. The form factor for mesons in $QCD_2$
has been derived in Ref. \cite{einhorn}
\begin{equation}
F^+\equiv \left(2p^+-q^+\right) F(q^2) 
= 2p^+(1-x)\left[f^{imp} + f^{vc} + f^{ec}\right],
\label{eq:f+}
\end{equation}
where $x=q^+/p^+$,
\begin{equation}
f^{imp} = \int_0^1 dz \phi\left(x+(1-x)z\right)\phi(z)
\label{eq:fimp}
\end{equation}
is the ``impulse term'' (Fig. \ref{fig:ffrho}a),
\begin{equation}
f^{vc} =-x^2 G^2 \int_0^1 du \int_0^1 dz \frac{\phi\left(x+(1-x)z\right)\phi(z)
G(u;q^2)}{\left[x(1-u)+(1-x)z\right]^2}
\label{eq:fvc}
\end{equation}
corresponds to a ``vertex-correction'' term (Fig. \ref{fig:ffrho}b), and
\begin{equation}
f^{ec} =x^2 G^2 \int_0^1 du \int_0^1 dz \frac{\phi(xu)\phi(z)
G(u;q^2)}{\left[x(1-u)+(1-x)z\right]^2}
\label{eq:fec}
\end{equation}
is the ``exchange current'' (Fig. \ref{fig:ffrho}c). $G(u;q^2)$ is the 
integrated Green's function in $QCD_2$
\begin{eqnarray}
G(u;q^2) &=& \int_0^1 dv G(u,v;q^2)\nonumber\\
G(u,v;q^2)&=&\sum_n \frac{\phi_n(u)\phi_n(v)}{q^2-\mu_n^2}.
\label{eq:green}
\end{eqnarray}
The fraction $x\equiv q^+/p^+$ of momentum transfer is determined by
energy conservation
\begin{equation}
\mu^2 = \frac{q^2}{x} + \frac{\mu^2}{1-x} .
\label{eq:econs}
\end{equation}
In the case of the $\pi$ form factor, the same diagrams contribute, but
there are two additional diagrams
which contain backward moving lines that connect to the meson via a
Gross-Neveu interaction (Fig. \ref{fig:ffpi})
The contribution of these diagrams to $F^+$ [Eq. (\ref{eq:f+})], 
the matrix element of $j^+$ is given by
\begin{equation}
Fig. \ref{fig:ffpi}a = 2p^+x^2Z\int_0^1 du \left[ \frac{M_{\bar{q}}}{1-xu}-
\frac{M_q}{x(1-u)}\right] \phi(xu) G(u;q^2)
\label{eq:ffpia}
\end{equation}
and
\begin{equation}
Fig. \ref{fig:ffpi}b = -2p^+x^2Z^2\int_0^1 du \frac{1}{1-xu} G(u;q^2),
\label{eq:ffpib}
\end{equation}
where $Z$ is a (dimensionful)
constant related to the normalization of the vertex. Its value
can be determined from
\begin{equation}
\phi(x)= -Z\int dy G(x,y,\mu^2) \left(\frac{M_q}{y}+\frac{M_{\bar{q}}}{1-y}
\right),
\end{equation}
where $\phi$ is normalized according to $\int_0^1dx\phi^2(x)=1$.
Note that we distinguished here between the masses of the quark and
anti-quark, which is not necessary for the $\pi$, but which will
be useful when we investigate $K$-mesons later.

In general, the wave functions and the integrals that determine the
form factor [Eqs. (\ref{eq:fimp})-(\ref{eq:ffpib})] have to be 
obtained numerically. However,
in the chiral limit the pion wave function is constant $\phi_\pi(x)=1$.
This yields $f^{imp}=1$. Furthermore, $f^{vc}$ and $f^{ec}$ exactly
cancel each other. Finally, with $M_q=M_{\bar{q}}=M$, we find $Z=M$ and there 
is also
a partial cancelation between the two diagrams in Fig. \ref{fig:ffpi},
yielding
\begin{eqnarray}
Fig. \ref{fig:ffpi} &=& -2p^+x\int_0^1 du
\frac{M^2}{1-u} G(u;q^2) \nonumber\\
&=&  p^+x\left[ 1-q^2\int_0^1du G(u;q^2)\right],
\label{eq:ffpi}
\end{eqnarray}
where we used that the solutions to 't Hooft's equation satisfy
(equal mass case)
$M^2\int_0^1 du \phi_n(u)/u(1-u)=\tilde{\mu}_n^2 \int_0^1 du \phi_n(u)$.
Adding up the various pieces and using furthermore [from Eq.
(\ref{eq:econs})] that $q^+=p^+$ (i.e. $x=1$) for vanishing $\pi$ mass, 
we thus find the remarkably simple result for the elastic vector form
factor of the pion
\begin{equation}
F_\pi(q^2) = 1-q^2\int_0^1 du G(u;q^2) =\sum_n \frac{\tilde{\mu}_n^2 
{g_n^V}^2}{\tilde{\mu}_n^2-q^2},
\label{eq:fpi}
\end{equation}
where $g_V(n)=\int_0^1du\phi_n(u)$.
In particular, we find for the rms-radius \footnote{Note that we use
here the 3+1 dimensional relation between the slope of the form factor and
the rms radius since this is supposed to be a model for 3+1 dimensional
QCD.}
\begin{equation}
\langle r^2 \rangle = -6\frac{d}{dq^2} F(q^2) = 6\sum_n\frac{g_V(n)^2}
{\tilde{\mu}_n^2},
\end{equation}
i.e. for the parameters used above one obtains a numerical value of
\begin{equation}
\sqrt{\langle r^2 \rangle_\pi} \approx 0.605 fm,
\end{equation}
which is only slightly smaller than the rms-radius for the $\rho$ meson
in the same model $\sqrt{\langle r^2 \rangle_\rho} \approx 0.625$.
The explanation for this near equality of rms radii is that both
form factors are dominated by the rho meson pole 
(note that $\sqrt{6}/m_\rho = 0.627fm$). \footnote{Of course,
for the $\rho$ meson it would be the $\omega$ meson pole, but in the
large $N_C$ limit the $\rho$ and $omega$ mesons are degenerate.}
The experimental value for the rms-radius of the pion is only slightly
larger $\sqrt{\langle r^2 \rangle_{\pi, exp}}=0.663 \pm 0.006 fm$. 
However, one should not be too impressed by this excellent agreement, since
the $\rho$ mass is an input parameter and since the rms radius of the pion
is dominated by the $\rho$ meson pole. The interesting aspect of the
model in this respect is that it reproduces vector meson dominance.

While the form factors of the $\pi$ and $\rho$ mesons are rather similar
for low momentum transfers, there is a qualitative difference between them
at large momentum transfers.
The elastic form factor of the $\rho$ meson follows the asymptotic behavior
of form factors in $QCD_2$, which is given by
\begin{equation}
F_\rho(Q^2)\sim \frac{1}{\left(Q^2\right)^{1+\beta}}
\quad
(Q^2 \rightarrow \infty),
\end{equation}
where $\beta$ is the same exponent that also appears in the end
point behavior of the wave function. For the parameters used in the
present fit ($m_q^2\approx G^2/\pi$) one finds $\beta \approx 0.5$ .  

Naively, from Eq. (\ref{eq:fpi}) one would expect $F_\pi(Q^2) \sim
1/Q^2$ as predicted by naive power counting. However,
$\sum_n \tilde{\mu}_n^2 g_P(n)g_V(n)$ diverges logarithmically and a more
careful analysis yields
\begin{equation}
F_\pi (Q^2) \longrightarrow 2M^2 \frac{\ln Q^2 }{Q^2} \quad
\quad (Q^2 \rightarrow \infty).
\label{eq:fpiasy}
\end{equation}
This logarithmic growth of $Q^2F_\pi (Q^2)$ arises because
we considered the limit of an infinitely heavy gluon in the collinear
QCD model. One should thus not take Eq. (\ref{eq:fpiasy}) literally
at very large $Q^2$, but only in an intermediate momentum range, where
it makes sense to introduce an infinitely heavy gluon as an effective degree
of freedom. 
\begin{figure}
\unitlength1.cm
\begin{picture}(15,7)(1.2,-9.5)
\includegraphics{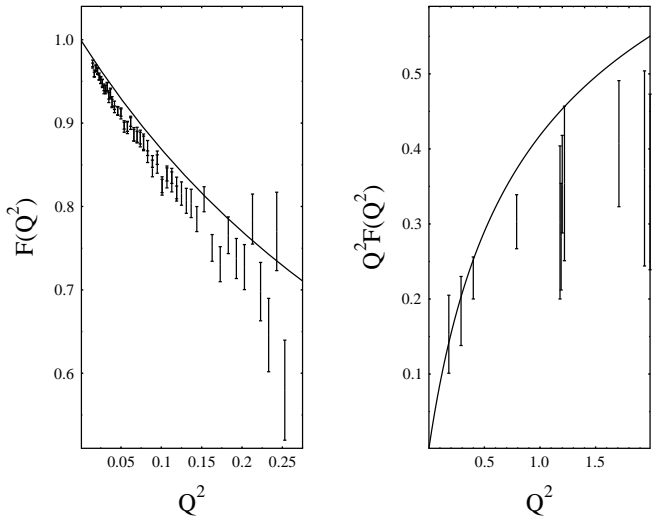}
\end{picture}
\caption{Numerical results for the pion form factor [Eq. (3.17)]. 
The experimental data is from Ref. [15]. 
}
\label{fig:formpi}
\end{figure}
While the details of the form factor analysis at large $Q^2$ are model
dependent, there are again some qualitative features which
might be important also in full $QCD_{3+1}$. Since there is an attractive
short range interaction acting only in the $\pi$ channel, the wave function
of the $\pi$ (in an equal time picture) contains much more high momentum
components than the $\rho$. As a result, the elastic form factor 
of the $\pi$ should fall off much slower than the form factor of the $\rho$.

\subsection{Strange Quarks}

The extension of the model to include strange quarks is straightforward and
any details will be omitted here. The only issue are the numerical values of
the model parameters, i.e. which of the parameters one should take from the
non-strange fit.

One might argue that the only new parameter is the strange quark mass, since
the gauge coupling is obviously flavor independent and since the Gross-Neveu
coupling also turns out to be flavor independent if one starts from a
flavor independent coupling of the quarks to the transverse component of the
gluons and if one strictly takes the limit of infinite mass for the  transverse
gluons. The results of a fit, where $m_s$ as the only new parameter is shown
in Table III. Using the $K$ mass as an input, one finds $m_s=502 MeV$, which is
quite heavy compared to a light quark mass of $m_q= 278 MeV$ (the fitted value
from the non-strange sector). As a result, the masses of the $K^*$ and the $\phi$
turn out to be much too high. Alternatively, if one uses the $K^*$ mass as as
input parameter, \footnote{Both the $K^*$ and the $\phi$ do not depend on the Gross-Neveu
coupling and are thus better suited for determining the strange quark mass
independently from the Gross-Neveu coupling.}
the $K$ turns out much too light.

However, one should keep in mind that one should consider the Gross-Neveu
interaction of the model an effective interaction, which is obtained by integrating
out high energy degrees of freedom. In real QCD, these degrees of freedom
do not have infinite mass and one would therefore expect some scale (and
thus also some flavor-) dependence of the effective interaction. Since we are
not able to perform this elimination procedure explicitly, we must also regard the
Gross-Neveu coupling for strange quarks as an independent parameter in this
phenomenological model. 
Corresponding results are shown in the forth column of Table III (fit 2).
The strange quark mass in this case (from fitting the
mass of the $K^+$) turns out to be $m_s=408 MeV$, which is much more reasonable
compared to $m_q$ than the above fit. The $K$ mass is fitted by construction
and yields the subtraction constant for the Gross-Neveu interaction, 
\footnote{In fit 1 in Table III, the numerical results for the $K$ and the
$K^*(1430)$ have large errors associated with them, since those calculations
do not subtract at the $K$-pole and one therefore has to deal numerically
with divergent quantities.}
which we used to calculate the mass of the lightest strange meson with $0^+$
quantum numbers. Compared with the lightest meson with these quantum numbers
that can be found in the particle data table, the model calculation gives a much
too small mass. This is to some extent surprising, since the model did so well
for the non-strange $0^+$ meson. However, we will not attempt to explain this
discrepancy here, since the quark model of these meson is to some extent still
very speculative \cite{swanson}.
\begin{table}
\label{tab:spectrum2}
\begin{tabular}{|c|l|l|l|l|}
meson & $J^P\quad\quad$ & fit 1 & fit 2 $\quad$& exp.(MeV)$\quad$\\ \hline
$K^+$ &      $0^-$  & $494 \pm 5$ &  494 & 494 \\
$K^*$ &      $1^-$  & 986  &  892 & 892 \\
$\phi$ &      $1^-$ & 1198 & 1012 & 1020 \\
$K^*(1430)$ & $0^+$ & $1180 \pm 10$ & 1129 & $ 1429 \pm 6$ 
\end{tabular}
\caption{Comparison of numerically calculated meson masses to experimental
data. In fit 1, all parameters other than the strange quark mass
are taken from the non-strange fit. The Kaon mass is taken as an input
parameter. In fit 2, the Gross-Neveu coupling is also allowed to differ from
the coupling in the non-strange sector.
}
\end{table}
The light-cone wave functions for mesons containing strange quarks are 
quite similar to wave functions of non-strange mesons and there is thus no
reason to display them here. The only difference is that the wave-functions
are no longer symmetric under $x\rightarrow 1-x$, but they are slightly
shifted such that the $s$ quark tends to carry more momentum than the light
quark, as one would naively expect.

With the high quality electron beam now available at Jefferson Laboratory,
there will be attempts to measure the electro-magnetic form factor
of the $K$-meson over a wide momentum range \cite{kaon}.

\begin{figure}
\unitlength1.cm
\begin{picture}(15,7)(1.2,-9.5)
\includegraphics{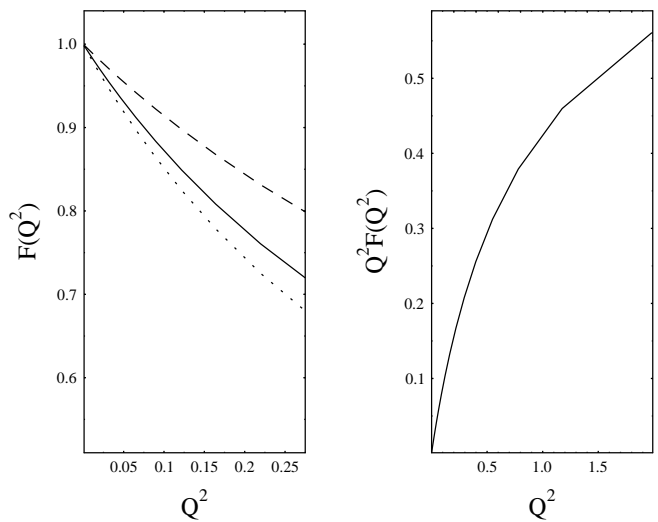}
\end{picture}
\caption{Numerical results for the kaon form factor. 
Dashed line: strange form factor, dotted line: up form factor, 
full line: electro-magnetic form factor. 
}
\label{fig:formk}
\end{figure}

In order to calculate this observable from our collinear QCD model
we concentrate on fit 2 (Gross-Neveu coupling adjusted), since it
is crucial to have both the correct $K$-meson mass as well as the
$\phi$-meson mass in order to obtain a reasonable model for the
Kaon form factor. Application of above formulas for the form factor
is tedious but straightforward and the result of such a calculation
is shown in Fig. \ref{fig:formk}. 
A comparison with Fig. \ref{fig:formpi} 
shows that the form factors of $\pi$ and $K$ mesons are rather similar.
At first, this seems surprising, but the physics of this result becomes
clear when one decomposes the form factor of the $K^+$ into the 
contributions from the $u$ and the $\bar{s}$ quark. 
The various form factors now find a simple interpretations in a quark model:
Clearly, the $u$ form factor falls off faster than the $s$ form factor,
since the heavier $s$ quark is more localized than the $u$ quark.
Furthermore, when comparing the $u$ form factor in a $K$ with the 
$u$ form factor in a $\pi$ we find that the $u$ quark is more localized
in the $\pi$ than in the $K$. There are two reasons for this trend.
First of all, the Gross-Neveu coupling is weaker in the $K$ and thus
there is a little less attraction in the $K$ compared to the $\pi$
(this compensates perhaps for the increased reduced mass).
Furthermore, while the mean separation between quark and anti-quark
changes perhaps only little when going from the $\pi$ to the $K$,
the center of mass gets shifted towards the $s$ quark, which means that 
the $u$ quark gets ``pushed out'' a little.
The combination of a slightly faster (compared to the $\pi$)
falloff of the $u$ form factor and the slower falloff of the
strange form factor (compared to the $\bar{d}$ in the $\pi$)
then results in a net electro-magnetic form factor for the $K$ meson
which is very similar to the $\pi$ form factor.

For the rms radii one finds 
$\sqrt{\langle r^2_s \rangle_K} \approx 0.490 \pm .005 fm$ 
and 
$\sqrt{\langle r^2_u \rangle_K} \approx 0.665 \pm .005 fm$ 
resulting in a net (electro-magnetic) rms radius of
$\sqrt{\langle r^2 \rangle_K} \approx 0.612 \pm .005 fm$
(experiment: 
$\sqrt{\langle r^2 \rangle_{K^+}} \approx 0.58 \pm .04 fm$ \cite{krms}), 
which is almost identical to the rms radius of the $\pi$ 
$\sqrt{\langle r^2 \rangle_\pi} \approx 0.605 fm$
in this model. 
For the neutral $K_0$, the model predicts
$\langle r^2 \rangle_{K^0} \approx = 0.067 \pm .005 fm^2$, which is
consistent with the experimental result 
$\langle r^2 \rangle_{K^0} \approx -0.054 \pm .026 fm^2$ \cite{krms}.

While numerical model results are consistent with available experimental
data \cite{krms}, the error bars of the latter are large and more
precise data, over a larger momentum range, would
be very useful \cite{kaon}.

\section{Summary}
We have considered the collinear model for QCD where all modes with nonzero
transverse momentum (relative to an arbitrary, but fixed, direction) are 
neglected. Such a model has not been derived from QCD, and should thus be 
regarded as a purely phenomenological model. Within this model, we focussed 
on the limit where the effective mass of the transverse
gluon becomes very large, but with an appropriately rescaled 
coupling to the quarks. As a result, one obtains a model that resembles
the 't Hooft model but with helicity degrees of freedom for the quarks and
point-like helicity-dependent quark-antiquark interactions that
resemble the chiral Gross-Neveu model. This limiting case of collinear 
QCD is solvable for large $N_C$.

Given the ad-hoc truncation of the degrees of freedom, one would be tempted
to dismiss such a model as not being a very useful model for $QCD_{3+1}$.
However, as the physical observables presented in this paper indicate,
the model yields a remarkably good description for many experimental data.

First of all, the spectrum of vector and axial-vector mesons turns
out to be rotational invariant (i.e. helicity independent !). 

Secondly,
with only three free parameters available [which are fixed to
reproduce the  $\pi$, $\rho$ and $\rho (1450)$ masses]
one obtains an extraordinarily good result for the masses of other 
non-strange mesons. 
One of the most surprising result in the calculations of the spectrum
was the excellent agreement with the mass of the $a_0(980)$, which is often
difficult in quark models and which is believed to possess a large
$K\bar{K}$ component (which is not there at large $N_C$).
We have no explanation for this surprisingly good agreement 
of the model results with the 
experimental data. This agreement might be accidental and it does
also depend on which physical mesons are identified with meson states
within the model. Since the model lacks transverse momenta, the model
contains of course much less states than for example the quark model
and therefore one must pick a subset of states with which the model
states are identified. While the choice made in the paper seemed by
far the most reasonable, there is some uncertainty in the assignment
of quantum numbers since the model lacks full rotational invariance.

The spectra of mesons containing strange quarks, were less well reproduced.
If one uses the same coupling in the strange and non-strange sectors,
then it turns out to be impossible to fit the $K$ and $K^*$ masses at the
same time. While the fit improves considerably, when one allows the Gross-Neveu
coupling for strange quarks to differ from the coupling for light quarks,
the overall fit for mesons with strange quarks never reaches the quality
of the fit for non-strange mesons. Specifically, the $J^{P}=0^{+}$
meson is not very well reproduced, which might be a hint that the agreement
in the non-strange sector is to some extent accidental. 

Motivated by the success of the meson spectrum calculations, we considered
other observables as well. An observable which is particularly easily 
accessible in the LF framework are the LF wave functions of hadrons.
The wave function for the $\pi$ turned out to be almost completely flat
[$\phi_\pi(x) =1$ in the chiral limit]. In contrast, the wave function
of the $\rho$ turned out to be strongly peaked in the middle, which is more 
reminiscent of a weakly bound state.
Physically, this difference did arise because the model has an attractive
zero-range interaction which acts only in the $\pi$ channel. A similar
picture arises when one considers the LF wave functions for the $a_0$ and
$b_1$ mesons. Both wave functions have nodes in the middle, which reflects 
the parity of these mesons, but only the wave function for the $a_0$
is non-vanishing at the boundary.

One feature which makes this model particularly useful for
form factor calculations is that it is
fully covariant under boosts in the longitudinal direction, which makes
the extraction of form-factors frame independent. 
Despite the differences in their LF wave functions, the $\rho$ and $\pi$
meson turn out to have almost the same rms radius (i.e. slope of the
vector form factor. The main reason for this result is vector meson 
dominance for the vector form factor at small momentum transfers, 
which is also a feature of 
collinear QCD. At large momentum transfers, the $\rho$ form factor falls
off much faster than the $\pi$ form factor. Roughly speaking, a flat LF wave 
function corresponds to an equal time wave function which has a very large
high momentum component (the end points of the LF wave function correspond to
infinite momenta for the constituents in an equal time framework) thus
resulting in less wave function suppression for the large momentum transfer
form factor.

In order to obtain a reasonable fit for the spectrum of mesons containing
strange quarks, it turns out to be necessary to re-fit the Gross-Neveu 
coupling.
After having done this one obtains an electro-magnetic form factor for
the $K$ meson, which is very similar to the form factor of the $\pi$.
This result is ``explained'' by $s$ form factor which falls of slower
and a $u$ form factor that falls of faster than $d$ and $u$ form factors
in the $\pi$ so that the net difference between $K$ and $\pi$ nearly vanishes.
The predicted charge radii are consistent with experimental data for $K^+$
and $K^0$ mesons.

\acknowledgements
The author thanks Craig Roberts, Eric Swanson and Adam Szczepaniak
for helpful comments and references. This work was supported by the D.O.E. 
(grant no. DE-FG03-96ER40965) and in part by TJNAF.
\appendix
\section{On the renormalization of the vertex mass}
\label{app:vertex}
As long as the same cutoffs are used in a covariant calculation and in a 
LF calculation \footnote{Under these circumstances, the only difference between
a covariant calculation and a LF calculation is that zero-modes are omitted in the
latter.}, the current quark mass of the covariant calculation and
the vertex mass of the LF calculation usually are the same. 
Chiral symmetry breaking manifests itself through the renormalization of the 
kinetic mass term in the LF calculation 
\cite{mb:parity,mb:hala,mb:iowa,mb:finite}.

The situation changes when the LF calculation is done using a Tamm-Dancoff 
approximation. This fact is best illustrated by considering a concrete example,
such as a fermion interacting with some boson field. For simplicity, let
us assume that the fermion self-energy has only a scalar piece and let us furthermore
assume that this scalar piece is momentum independent \footnote{An explicit model
with such features is provided by a model where the boson mass goes to infinity
and where the coupling is rescaled such that a nontrivial mass renormalization
remains.} --- in other words, let us assume that
\begin{equation}
\Sigma (p) = \delta m ,
\end{equation}
where $\delta m$ is a constant. 

\begin{figure}
\unitlength1.cm
\begin{picture}(15,9)(2,1)
\includegraphics{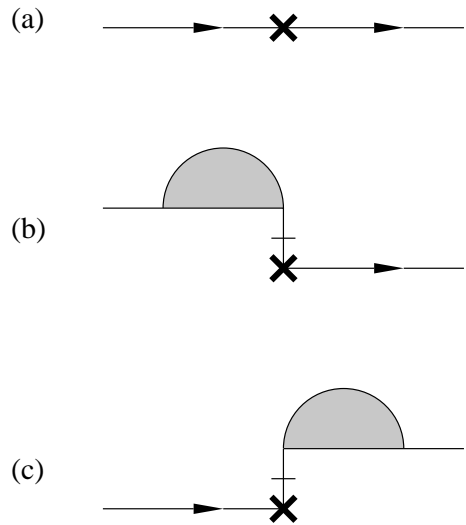}
\end{picture}
\caption{Diagrams contributing to the three-point vertex for a physical fermion.
(a) bare vertex, (b) self-energy insertion on the outgoing fermion line,
(c) self energy insertion on the incoming fermion line.
}
\label{fig:vertex}
\end{figure}

Without truncations of the Fock space, the three point interaction for
the physical fermion (=solution to the Hamiltonian) gets modified through 
Feynman diagrams
that include self-energies connected to the vertex by an instantaneous 
fermion line (Fig. \ref{fig:vertex}).
Note that under the above assumptions about the self-energy $\Sigma$, there are no 
further diagrams contributing to the physical vertex.

For the case of a (transverse) vector vertex \footnote{Other cases are analogous.},
the bare vertex yields a matrix element for helicity flip transitions
\begin{equation}
\Gamma_{bare}^{flip} = \frac{m^0_V}{\sqrt{q^+}}\left( \frac{1}{p^+} - \frac{1}{{p^\prime}^+}
\right),
\end{equation}
where $q^+\equiv p^+ - {p^\prime}^+$ is the momentum transfer at the vertex.
The two diagrams with a self energy insertion in the incoming and outgoing
external lines yield, respectively
\begin{eqnarray}
\delta \Gamma_{in}^{flip}&=&\frac{\delta m}{\sqrt{q^+}}\frac{1}{p^+}
\nonumber\\
\delta \Gamma_{out}^{flip}&=&-\frac{\delta m}{\sqrt{q^+}}
\frac{1}{{p^\prime}^+},
\end{eqnarray}
so that
\begin{eqnarray}
\Gamma^{flip}&\equiv& \Gamma_{bare}^{flip}+ 
\delta \Gamma_{in}^{flip}+ \delta \Gamma_{out}^{flip}
\nonumber\\
&=& \frac{m^0_V+\delta m}{\sqrt{q^+}}\left( \frac{1}{p^+} - \frac{1}{{p^\prime}^+}
\right) .
\end{eqnarray}
This example shows that, if one imposes a Tamm-Dancoff (TD) truncation
such that the $\delta m$ correction to the vertex can no longer be 
generated through higher order Fock components, then one needs to
renormalize the vertex mass accordingly in order to obtain the
same physical results as without the TD truncation.
Hence, with a TD approximation in place, the numerical value
of the vertex mass need no longer agree with the numerical value
of the current mass. In particular with a TD approximation, there
is no reason that the vertex mass should vanish in the chiral limit.
This is in contradistinction to LF calculations without TD 
approximations \cite{mb:hala}.

\section{The 't Hooft -- Gross -- Neveu Model}
\label{sec:hgn}
Because the Gross-Neveu interaction is  quadratic in the
non-dynamical component of the fermion field $\psi_-$, the
constraint equation becomes non-linear and a canonical
LF formulation of models with such an interaction is somewhere
between difficult and impossible. Usually, one has to 
use some indirect methods for solution. In this Appendix,
we will use covariant methods, supplemented by the LF
solution to $QCD_{1+1}$, to construct a solution
to the Gross-Neveu model with additional
$QCD_{1+1}$ interactions.

Let us consider the 1+1 dimensional model
\begin{eqnarray}
{\cal L} &=& \bar{\psi}\left( i\partial \!\!\!\!\!\!\not\,\; -
\frac{G}{\sqrt{N}}A\!\!\!\!\!\!\not\;\,-m\right)\psi
- \frac{1}{2}\mbox{tr}\left(F_{\mu \nu}F^{\mu \nu}\right)
\nonumber\\
& &-\frac{\gamma}{2N}\left[ \left(\bar{\psi}\psi\right)^2
-\left(\bar{\psi}i\gamma_5\psi\right)^2\right],
\label{eq:hgn}
\end{eqnarray}
where $\psi$ has $N$ color components (fundamental representation),
$F_{\mu \nu}=\partial_\mu A_\nu
-\partial_\nu A_\mu +i\frac{G}{\sqrt{N}} \left[A_\mu,A_\nu\right]$ 
is the gauge field and the limit $N\rightarrow \infty$ is implied.
Obviously, for $\gamma =0$ one obtains the 't Hooft model
($QCD_{1+1}$ in the limit $N\rightarrow \infty$) \cite{thooft}, 
while for $G=0$ the model
reduces to the chiral Gross-Neveu model \cite{gn}.
Because of the four fermion interaction, ${\cal L}$ has to be
renormalized. \footnote{We closely follow Ref. \cite{brian},
where more details can be found.} 

Before we derive the bound state equation corresponding to
Eq. (\ref{eq:hgn}), let us illustrate the techniques in the
simpler case $G=0$ (the chiral Gross-Neveu model): First of
all, the scalar tadpole gives rise to a ``constituent mass''
$M$ for the fermion. Secondly, since we are considering the
$N\rightarrow \infty$ limit, the meson spectrum can be determined
by summing up a (geometric) series of bubbles, yielding
for the pseudo-scalar two-point function
\footnote{Up to an overall normalization factor, which is not relevant for
determining bound state masses.}
\begin{equation}
D_\pi(p^2)=\left(\frac{1}{\gamma} -i
\int \frac{d^2q}{(2\pi)^2} \mbox{tr}
\frac{1}{\left(q\!\!\!\!\!\not\,\,-M\right)
\left(q\!\!\!\!\!\not\,\,-p\!\!\!\!\!\not\,\,-M\right)}
\right)^{-1} .
\label{eq:ps2point}
\end{equation}
The integral in Eq. (\ref{eq:ps2point}) diverges logarithmically.
We renormalize by subtracting at $p^2=0$ (strictly speaking
we first introduce a regulator, then subtract and then send
the regulator to infinity), yielding
\begin{equation}
D_\pi(p^2)=\left(\frac{1}{\gamma_{ren}} -
\frac{1}{2\pi}\int_0^1 dx 
\frac{p^2}{M^2-p^2x(1-x)}\right)^{-1} ,
\label{eq:ps2point2}
\end{equation}
where the renormalized (finite!) coupling is related
to the bare coupling (zero!) via (cut-offs not explicitly shown)
\begin{eqnarray}
\frac{1}{\gamma_{ren}} &=&\frac{1}{\gamma}
 -i
\int \frac{d^2q}{(2\pi)^2} \mbox{tr}
\frac{1}{\left(q\!\!\!\!\!\not\,\,-M\right)
\left(q\!\!\!\!\!\not\,\,-M\right)}
\nonumber\\
&=&\frac{1}{\gamma}- \frac{1}{2\pi}\int_0^1 \frac{dx}{x(1-x)} .
\label{eq:lren}
\end{eqnarray}
The pseudo-scalar bound state (in the following referred to as
the pion) mass is obtained in terms of the 
renormalized coupling be searching the pole of $D_\pi(p^2)$.
In practice, one often turns the argument around and uses the pion
mass as an input parameter, which yields the renormalized coupling as
a function of the pion mass
\begin{equation}
0\stackrel{!}{=}\frac{1}{\gamma_{ren}} -
\frac{1}{2\pi}\int_0^1 dx 
\frac{\mu_\pi^2}{M^2-\mu_\pi^2x(1-x)} .
\label{eq:ps2point2b}
\end{equation}
Since the 't Hooft model is most conveniently solved in the 
LF framework, we will also formulate  the 't Hooft -- Gross -- Neveu
model using LF quantization. As a warmup exercise, we thus
reconsider the Gross-Neveu model in this framework as well. 
Using standard canonical quantization plus renormalization of
the kinetic mass plus (as explained in Appendix \ref{app:vertex})
renormalization of the vertex mass
one finds the (still ill defined!) bound state equation for 
pseudo-scalar mesons
\begin{equation}
\mu^2 \psi(x) = \frac{M^2 \psi(x)}{x(1-x)}
- \frac{\gamma M^2}{x(1-x)} \int \frac{dy}{2\pi} 
\frac{\psi(y)}{y(1-y)}
\label{eq:gnlf1}
\end{equation}
(for scalar mesons, where the wave-function is odd under $x\rightarrow
1-x$, another term contributes which is omitted here for simplicity).
Note that we set the vertex mass equal to the kinetic mass
in Eq. (\ref{eq:gnlf1}). As is explained in Appendix 
\ref{app:vertex}, the vertex mass can acquire nontrivial 
renormalization, if certain corrections are suppressed by
the approximations used
(here: fermion tadpoles connected to the bare vertex by an 
instantaneous fermion line). The fact that we set it here equal
to the kinetic mass term is arbitrary and irrelevant, since
the interaction term is also multiplied by the (still undetermined)
bare coupling constant. But, in order to achieve maximum similarity
with the covariant approach, we set the vertex mass equal to
the kinetic mass at this point.

Eq. (\ref{eq:gnlf1}) yields a divergent solution, since
$\psi (x) \stackrel{ x\rightarrow 0,1}{\longrightarrow} const.$
and thus $\int dy \psi(y)/y(1-y)$ diverges.
This divergence is exactly the same divergence that one obtains
in the covariant approach as well (before coupling constant
renormalization) as one can see by reducing Eq. (\ref{eq:ps2point})
to a parameter integral, yielding
\begin{equation}  
D_\pi(p^2)=\left(\frac{1}{\gamma}
-\frac{1}{2\pi}\int_0^1 \frac{dx}{x(1-x)} 
\frac{M^2}{M^2-p^2x(1-x)}\right)^{-1}
\label{eq:ps2point3}
\end{equation}
and the whole divergence can be avoided provided one renormalizes 
properly. Here we will use a trick 
\footnote{A similar trick has been used by T. Heinzl (talk given
at the ``1997 workshop on LF quantization'' in Les Houches, France).}
to arrive at a renormalized bound
state equation: the
operator identity
\begin{equation}
\partial_\mu \left( \bar{\psi} \gamma^\mu \gamma_5
\psi \right) = 2m\bar{\psi}i\gamma_5 \psi,
\end{equation}
where $m$ is the current quark mass, together with Lorentz covariance,
implies for vacuum to meson (on shell) matrix elements
\begin{equation}
p^-\langle 0| \bar{\psi} \gamma^+ \gamma_5\psi| n, p \rangle
= m \langle 0| \bar{\psi} \gamma_5 \psi | n,p \rangle ,
\end{equation}
i.e. in terms of the wave-function
\begin{equation}
\mu_n^2 \int_0^1 dx \psi_n(x) = m  \int_0^1 dx\frac{M}{x(1-x)}\psi_n(x) 
\end{equation}
The reason $M$ and not $m$ appears in the LF expression for 
$\langle 0| \bar{\psi} i\gamma_5 \psi | \pi (p) \rangle $ is
again the Tamm-Dancoff expansion used.
We can use this identity to express the divergent integral in Eq. 
(\ref{eq:gnlf1}) by a convergent integral, yielding the renormalized
(but non-hermitian, since the Hamiltonian depends on the
eigenvalue) bound state equation for the pion channel
\begin{equation}
\mu^2 \psi(x) = \frac{M^2 \psi(x)}{x(1-x)}
- \frac{\gamma M\mu^2}{m x(1-x)} \int \frac{dy}{2\pi} \psi(y) .
\label{eq:gnlf2}
\end{equation}
This ``miracle'' is possible since the current quark mass itself is cutoff
dependent and goes to zero (for fixed physical masses) as the cutoff is
sent to infinity. In fact, it goes to zero in such a way that the
ratio $\gamma/m$ stays finite, which motivates the definition
\begin{equation}
\gamma_{ren}^{LF} \equiv \frac{\gamma M}{m} .
\end{equation}
A consistency check of this result is obtained by explicitly solving
the renormalized integral equation. From Eq. (\ref{eq:gnlf2}) one finds
\begin{equation}
\psi(x) = \frac{\gamma_{ren}^{LF} \mu^2}{ M^2-x(1-x)\mu^2 }\int \frac{dy}{2\pi} \psi(y).
\end{equation}
Consistency requires that
\begin{equation}
1 \stackrel{!}{=} \gamma_{ren}^{LF} \mu^2 \int_0^1\frac{dx}{2\pi} 
\frac{1}{ M^2-x(1-x)\mu^2 },
\end{equation}
which we recognize as the pole condition for $D_\pi(p^2)$ 
[Eq. (\ref{eq:ps2point2})]
with $\gamma_{ren}$ replaced by $\gamma_{ren}^{LF}$.

Let us now proceed to the full 't Hooft -- Gross -Neveu model
(\ref{eq:hgn}). A calculation analogous to the covariant 
procedure discussed above can again be obtained by summing up
a chain of pseudo-scalar bubbles, but now each bubbles is
dressed by the $QCD_{1+1}$ interactions as well as by the
Gross -- Neveu (tadpole) self-energy. \footnote{The fact that
the fermion lines within the Gross -- Neveu tadpoles are dressed
by $QCD_{1+1}$ interactions only changes the self-mass by a 
finite amount. Since the induced mass is only a bare parameter anyway, 
this additional renormalization of a parameter that is already
renormalized has no further consequences.}
\begin{figure}
\unitlength1.cm
\begin{picture}(15,4)(0.3,-5)
\includegraphics{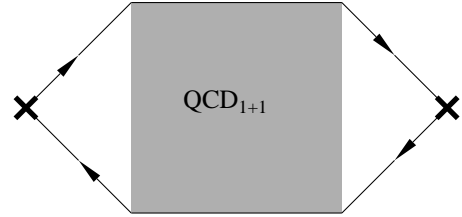}
\end{picture}
\caption{In the pseudo-scalar two-point function, the $q\bar{q}$
propagation within each bubble is governed by the $QCD_{1+1}$
Kernel in planar approximation ($\rightarrow$ 't Hooft model).
}
\label{fig:bubble}
\end{figure}
In practice, these
bubbles can be evaluated by inserting a complete set of 
('t Hooft-) meson states
\begin{equation}
D^{-1}_\pi (p^2) = \frac{1}{\gamma} -\frac{1}{2\pi}
\sum_n \frac{g_P(n)^2}{\tilde{\mu}_n^2-p^2},
\label{eq:hgn1}
\end{equation}
where the $\tilde{\mu}_n$ and $g_P(n)$ are the masses and pseudo-scalar
couplings of the n-th meson solution for the 't Hooft equation.
The masses are most conveniently calculated in terms of the
LF wave-functions
\begin{equation}
g_P(n)= M\int_0^1 dx \left(\frac{1}{x}+\frac{1}{1-x}\right) \psi_n(x)
.
\label{eq:gpn}
\end{equation}
The sum in Eq. (\ref{eq:hgn1}) diverges, but since $QCD_{1+1}$
is superrenormalizable, this is only the free field divergence.
We renormalize again by performing a zero momentum subtraction,
yielding
\begin{equation}
D^{-1}_\pi (p^2) = \frac{1}{\gamma_{ren}} -\frac{1}{2\pi}
\sum_n \frac{p^2 g_P(n)^2}{\tilde{\mu}_n^2
\left(\tilde{\mu}_n^2-p^2\right)},
\label{eq:hgn2}
\end{equation}
where
\begin{equation}
\frac{1}{\gamma_{ren}} =\frac{1}{\gamma}
- \sum_n \frac{ g_P(n)^2}{\tilde{\mu}_n^2}
\label{eq:hgn3}
\end{equation}
is finite. For the scalar channel we are {\it not} free to renormalize
independently, since chiral symmetry demands that the scalar and
pseudo-scalar couplings are the same. This yields
\begin{eqnarray}
D^{-1}_\sigma (p^2) &=& \frac{1}{\gamma}-\frac{1}{2\pi}
\sum_n \frac{g_S(n)^2}{\tilde{\mu}_n^2-p^2}
\nonumber\\
&=&\frac{1}{\gamma^S_{ren}} -\frac{1}{2\pi}
\sum_n \frac{p^2 g_S(n)^2}{\tilde{\mu}_n^2
\left(\tilde{\mu}_n^2-p^2\right)},
\label{eq:hgn4}
\end{eqnarray}
where 
\begin{equation}
g_S(n)= M\int_0^1 dx \left(\frac{1}{x}-\frac{1}{1-x}\right) \psi_n(x)
\label{eq:gsn}
\end{equation}
are the scalar couplings and
\begin{equation}
\frac{1}{\gamma^S_{ren}} = \frac{1}{\gamma^P_{ren}}
+ \frac{1}{2\pi}\sum_n \frac{g_S(n)^2- g_P(n)^2}{\tilde{\mu}_n^2}
\end{equation}
accounts for the (finite) differences in the zero momentum
subtraction in the scalar and pseudo-scalar channels.
For practical calculations, the following representation is 
also useful
\begin{equation}
D^{-1}_\sigma (p^2)=D^{-1}_\pi (p^2) + \frac{1}{2\pi}\sum_n \frac{g_S(n)^2- g_P(n)^2}{\mu_n^2}
\label{eq:hgn4b}
\end{equation}
Eqs. (\ref{eq:hgn2}) and (\ref{eq:hgn4}) or (\ref{eq:hgn4b}) are perfectly suitable for
a numerical determination of the spectrum of the
't Hooft -- Gross -- Neveu model. However, since we want
to determine LF wave-functions and parton distribution functions,
we also need to understand how to renormalize the LF wave equation
for this model. The canonical procedure (modulo renormalization
of both kinetic and vertex masses as explained above) yields
the (still ill defined) LF bound state equation for this model
\begin{eqnarray}
\mu^2 \psi(x) &=& \frac{M^2 \psi(x)}{x(1-x)}
+\int_0^1 dy \frac{ \psi (x)-\psi(y)}{(x-y)^2}
\nonumber\\
&-& \frac{\gamma M^2}{x(1-x)} 
\int_0^1 \frac{dy}{2\pi} \frac{\psi(y)}{y(1-y)}
\nonumber\\
&-& \frac{\gamma M^2(1-2x)}{x(1-x)} 
\int_0^1 \frac{dy}{2\pi} \frac{(1-2y)\psi(y)}{y(1-y)}.
\label{eq:hgnlf1} 
\end{eqnarray}
In the pseudo-scalar channel, we can use the same procedure that 
we introduced in connection with the Gross -- Neveu model, namely
replacing the pseudo-scalar coupling by the vector coupling, which yields
the renormalized LF bound state equation for the 'Hooft --
Gross -- Neveu model in the
pseudo-scalar channel (i.e. only even wave functions)
\begin{eqnarray}
\mu^2 \psi(x) &=& \frac{M^2 \psi(x)}{x(1-x)}
+G^2\int_0^1 dy \frac{ \psi (x)-\psi(y)}{(x-y)^2}
\nonumber\\
&-& \frac{\gamma_{ren}^{LF} M^2\mu^2}{x(1-x)} 
\int_0^1 \frac{dy}{2\pi}\psi(y).
\label{eq:hgnlf2} 
\end{eqnarray}
It is quite easy to verify consistency of the renormalized
LF equation (\ref{eq:hgnlf2}) with the more conventionally
obtained result (\ref{eq:hgn2}). Using the Green's function
for the 't Hooft equation,
\begin{equation}
G(x,y,p^2) \equiv \sum_n \frac{\psi_n(x)\psi_n(y)}
{\tilde{\mu}_n^2-p^2},
\end{equation}
one can rewrite Eq. (\ref{eq:hgnlf2}) in the form
\begin{equation}
\psi(x) = \int_0^1 dy G(x,y,\mu^2)
\frac{\gamma_{ren}^{LF} M^2}{y(1-y)} \int_0^1 \frac{dz}{2\pi}\psi(z).
\label{eq:hgnlf3}  
\end{equation}
Upon integrating over $x$ and using the fact that the solutions
to 't Hooft's equation satisfy
\begin{equation}
\tilde{\mu}_n^2 \int_0^1 dx \psi_n(x) = M^2 \int_0^1 dx 
\frac{\psi_n(x)}{x(1-x)}
\end{equation}
we recover Eq. (\ref{eq:hgn2}) --- provided we identify
$\gamma_{ren}^{LF}$ with $\gamma_{ren}$.

For scalar mesons, we have not been able to derive a renormalized LF 
bound state equation. However, this does not prevent us from calculating
the LF wave functions of scalar resonances, using the following trick: 
\footnote{We demonstrate this trick only for scalar meson but it works 
analogously for pseudo-scalar mesons.} From Eq. (\ref{eq:hgnlf1}) one notes
that the wave functions for scalar states satisfy
\begin{equation}
\psi_n(x) \propto \int_0^1 dy \, G(x,y,\mu_n^2)
\left(\frac{M}{x}-\frac{M}{1-x}\right),
\label{eq:trick}
\end{equation}
where $M_n^2$ is the invariant mass of the bound state. Thus one can first
obtain the bound state mass from the poles of $D_\sigma(p^2)$ [i.e. from Eq.
(\ref{eq:hgn4b})] and then one determines
$\psi_n(x)$ (up to a normalization factor) by solving the linear equation
corresponding to Eq. (\ref{eq:trick}), i.e. one solves
\begin{equation}
\left( \mu_\sigma^2 -H_{th}\right) \psi = M\left(\frac{1}{x}-\frac{1}{1-x}
\right)
\end{equation}
for $\psi (x)$, where $H_{th}$ is the `t Hooft Hamiltonian.

\end{document}